\documentclass[12pt,letterpaper,superscriptaddress]{JHEP3}
\pdfoutput=1
\usepackage{amssymb,amsmath,amsfonts,amstext,graphics}
\usepackage{cite}
\usepackage{graphicx}
\usepackage{epstopdf}
\input epsf.tex

\newcommand{\gsim}{\lower.7ex\hbox{$\;\stackrel{\textstyle>}{\sim}\;$}}
\newcommand{\lsim}{\lower.7ex\hbox{$\;\stackrel{\textstyle<}{\sim}\;$}}
\def\OO{{\cal O}}
\def\LL{{\cal L}}

\def\eg{{\it e.g.}}

\newcommand{\TeV}{\,\mathrm{TeV}}
\newcommand{\GeV}{\,\mathrm{GeV}}

\newcommand{\fb}{\,\mathrm{fb}}
\newcommand{\ifb}{\,\mathrm{fb}^{-1}}

\newcommand{\eff}{{\text{eff}}}

\newcommand{\MET}{\mbox{$E_T\hspace{-0.252in}\not\hspace{0.22in}$}}

\newcommand{\DO}{{\mbox{D0\hspace{-0.100in}$\not$\hspace{0.13in}}}}

\newcommand{\bef}{\begin{figure}[htbp]\begin{center}}
\newcommand{\eef}{\end{center}\end{figure}}

\newcommand{\arts}{\alpha_{\text{RTS}}}

\title{
\begin{flushright}
\mbox{\normalsize SLAC-PUB-13804}
\end{flushright}
\vskip 15 pt
Bigger, Better, Faster, More  at the LHC}
\author{Eder Izaguirre$^1$,
Michael Manhart$^2$, and 
Jay G. Wacker\,$^1$\\
$^1$Theory Group, SLAC,  Menlo Park, CA 94025\\
$^2$Department of Physics and Astronomy, Rutgers University,
Piscataway, NJ 08854
}

\abstract{
Multijet plus missing energy searches provide universal coverage for
theories that have new colored particles that decay into a dark matter candidate and jets.
These signals appear at the LHC further out on the missing energy tail than two-to-two scattering indicates.  
The simplicity of the searches at the LHC contrasts sharply with the Tevatron where more elaborate searches are necessary to separate signal from background.
The searches presented in this article effectively distinguish signal from background for any  theory where the LSP is a daughter or granddaughter of the pair-produced  colored parent particle without ever having to consider missing energies less than 400 GeV.  
}

\begin{document}

\section{Introduction}
\label{Sec: Introduction}
A new energy frontier is being explored with the commissioning of the LHC, allowing for  the best chance to discover physics beyond the Standard Model in over a decade.  One of the most robust signals of new physics is anomalous contributions to jets plus missing energy from new strongly interacting particles that subsequently decay into jets and an invisible particle that may be the dark matter. 

In advance of the LHC, there have been many studies of the discovery potential in jets and missing energy  \cite{SusyBenchmarks,JetsMETSearches,UEDPheno}.  Most of these studies study how to discovery a specific theory, \eg the mSUGRA parameterization of the minimal supersymmetric Standard Model (MSSM)  \cite{CMSSM} (see \cite{AMSB,GaugeMediation, MirageMediation,MiscSusy,Others} for other specific supersymmetric spectra), Universal Extra Dimensions (UED)s  \cite{UED} or Little Higgs theories \cite{LH}.  These searches present generally optimistic results for each specific theory, but different theories, even within the SSM may exhibit different signatures, \cite{Berger:2008cq}. How broadly applicable the searches optimized for specific theories are is an open question. This article begins to answer this question by studying which searches are necessary to provide universal coverage for a broad class of spectra.  

   
 Model-independent searches at hadron colliders are experimentally attainable by focusing on deviations from the Standard Model, and by parametrizing the details of a model in term of relevant observables like the masses of the particles produced \cite{ModelIndependent}.  \cite{GeneralizedJetsMET,ModelIndependentJetsMET} constructed a general search strategy at the Tevatron for discovering new physics in jets and missing energy  ($\MET$) regardless of the spectrum of new physics.  
 The search strategy required combining several different search channels:  $1j+\MET$, $2j+\MET$, $3j+\MET$ and $4^+j+\MET$, where the last channel was inclusive.   Inside each of these search channels, it was necessary to perform many measurements varying the visible energy ($H_T$) and missing energy.   Several dozen measurements were necessary to fully cover the Tevatron's discovery potential.    The primary challenge for the Tevatron is that new particles that the Tevatron can produce have similar masses to the Standard Model background processes.  At the LHC, the new particles that can be searched for are typically much heavier than the SM background processes (dominantly top pair production) and therefore are more energetic.  If there are new low mass colored particles that are similar in mass to the top, they are produced in such copious quantities that they are difficult to miss.
This difference between the Tevatron and the LHC results in a much simpler search strategy and the possibility of a missed discovery is greatly reduced.
    

The search strategy presented in this article will provide coverage for nearly all new colored particles that can be discovered in jets plus missing energy searches.  
This search strategy only uses the multijet ($4^+j+ \MET$) channel
; the efficacy of this strategy originates from highly energetic initial state radiation (ISR) and final state radiation (FSR) jets that accompany
new colored particle production at $\sqrt{s}=14\TeV$ at the LHC \cite{Alwall:2009zu}.  
Parton shower/matrix element (PS/ME) \cite{PSME} matching is used on both the {\it background and signal} in order to more accurately estimate the spectrum of additional hard radiation\cite{SUSYPSME}.  PS/ME matching increases the jet multiplicity of the signal, but it also increases the missing energy in signal events and therefore the discovery potential of the LHC.

The efficacy of this search strategy is demonstrated through three benchmark supersymmetric field contents:
\begin{itemize}
\item
Pair-produced gluinos in the supersymmetric Standard Model (SSM). Each gluino directly decays to two  light-flavored quarks and the LSP.   The LSP is the daughter of the original pair produced particle.
\item Pair-produced gluinos that cascade decay via a wino  and jets with the wino subsequently decaying to a $W^\pm$ and the LSP.   The LSP is the granddaughter of the original pair-produced particle. 
\item
 Pair-produced  squarks\footnote{The 9 flavors of squarks  $(\tilde{q}_i, \tilde{u}_i^c, \tilde{d}^c_i)$ are degenerate. } in the SSM that  decay to a quark and the LSP.
\end{itemize}
These benchmark models are illustrative of the wider class of theories that produce jets plus missing energy signatures beyond supersymmetric and apply to UEDs and other theories which have new colored particles that decay into a dark matter candidate.  The search strategies presented in this article are equally effective at separating gluinos that decay into multiple hard  jets plus missing energy from background as they are at separating squarks that decay into a few  soft jets plus missing energy from background.   

The organization of the article is as follows.
Sec. \ref{Sec: Strategy} presents the general strategy for model independent jets plus missing energy searches.  Sec. \ref{Sec: Generation} describes the methods used to calculate the SM backgrounds and new physics signals.
Sec. \ref{Sec: Results} shows the application of the search strategies to several minimal models.
Finally, Sec. \ref{Sec: Discussion} discusses the types of physics that are not effectively searched for with the
strategies described in this article.

\section{Search Strategy }
\label{Sec: Strategy}

The goal of this article is to demonstrate that a simple search strategy will discover 
a broad class of new physics models without any loss of sensitivity to unexpected theories.
 In \cite{ModelIndependentJetsMET}, the searches are first divided into mutually exclusive event samples with different numbers of jets:
\begin{eqnarray}
1j+\MET \quad 2j+\MET\quad 3j+\MET \quad 4^+j+\MET
\end{eqnarray}
where $4^+j+\MET$ is an inclusive, four or more jets plus missing energy search.  At the Tevatron it
is necessary to perform a series of searches in each of the multiplicities above in order to 
maximize the discovery potential for the full range of potential signals.   

This article will demonstrate that a series of measurements using only the $4^+j+\MET$ channel is sufficient to gain broad sensitivity to essentially all theories that are visible in jets plus missing energy searches at the LHC.  
The effectiveness of the multijet search, even in theories with $m_{\tilde{g}}\simeq m_{\chi^0}$, is a remarkable feature of a $\sqrt{s}=14\TeV$ LHC and requires that initial and final state radiation produce many, and occasionally all, of the jets used in the multijet search. 

The origin of this universality of the multijet search arises from several factors.
The first is that radiation allows  access to different components of the parton distribution functions (pdfs).  Pair-produced new particles originate from the $gg$ or $q\bar{q}$ initial states; however, an additional initial state jet allows access to the $qg$ pdf  that dominates over $gg$ above $\sqrt{\hat{s}} \simeq 700 \GeV$.    With two additional final state jets, pair production has access to the $qq$ pdf and dominates above $\sqrt{\hat{s}}\simeq 2\TeV$.    Therefore, including additional jets in the signal allows access 
to more energetic pdfs which in turn allows for more energetic
events.

For compressed spectra, there is a kinematic reason why addition radiation is so important for distinguishing signal from background. Consider the case where a jet with energy $E_j$ is radiated off a massive particle, $X$,  of mass $m$. The four momenta of the jet and the massive particle are given as
\begin{eqnarray}
p^\mu_j  = (E_j , E_j )\qquad
p^\mu_X  = \left(\sqrt{m^2 + E^{2}_j}, - E_j\right).
\end{eqnarray}
The additional cost of energy for radiating a jet off a massive particle is given by
\begin{eqnarray}
\delta\sqrt{\hat{s}}_S  =  & E_j + \sqrt{m^2 + E^{2}_j} - m  = & 
\begin{cases}  
E_j & \text{$m\gg E_j$}\\
2 E_j & \text{$m\ll E_j$}
\end{cases} .
\label{Eq: EnergyCostSignal}
\end{eqnarray}
Conversely, when radiating a jet off the SM backgrounds, letting $m\rightarrow\sqrt{s}$ in Eq.~\ref{Eq: EnergyCostSignal}, with $E_j \simeq 2 m_{\tilde{g}} \gg \sqrt{s}_B$ gives
\begin{eqnarray}
\delta\sqrt{\hat{s}}_B  =  E_j + \sqrt{\hat{s}_B + E^{2}_j} - \sqrt{\hat{s}}_B  = 
2 E_j .
\label{Eq: EnergyCostBackground}
\end{eqnarray}
Comparing signal and backgrounds yields
\begin{eqnarray}
\sqrt{\hat{s}}_{S+j} \simeq\sqrt{\hat{s}}_S + E_j & \qquad& \sqrt{\hat{s}}_{B+j}  \simeq  2E_j + \frac{\hat{s}_B}{2E_j} ;
\end{eqnarray}
letting $\sqrt{\hat{s}}_{S+j}\simeq \sqrt{\hat{s}}_{B+j}$ equalizes the CM energy for signal and background, thus increasing $S/B$. A similar procedure leads to the same relation in Eq.~(2.5) when the jet comes from initial state radiation.
The multitude of jets greatly simplifies the search strategy and reduces the number of measurements necessary to gain broad coverage of theories  beyond the Standard Model.

The remaining portion of this section is organized as follows.  Sec. \ref{Sec: Search Channels} classifies the search
channels into different jet multiplicities and addresses the cuts on jet $p_T$s.   Sec. \ref{Sec: Observables} introduces
the other kinematic and event shape variables that are useful at separating signal from background and specifies the series of searches that comprise the full search strategy.

\subsection{Classification of Search Channels}
\label{Sec: Search Channels}


In order to both trigger and ensure that the events are sufficiently distinctive compared to poorly understood backgrounds, the searches considered in this article have tighter cuts than those considered in the published search strategies from ATLAS \cite{ATLAS_NOTES} and CMS \cite{CMS_NOTES1}.  
The cuts used in this article  are loosely based on the benchmark ATLAS multijet + $\MET$ searches and the modifications to the searches tighten the searches\cite{Akimoto:2009zz}.    This article finds that requiring a larger  missing energy criteria is more effective  at  separating signal from background. 
 
 The first step in devising a model-independent, inclusive search strategy is to classify events into different jet multiplicities.  The primary variables used are the $p_T$s of the jets.   Each multiplicity has a definition of a high $p_T$, central primary jet.   After the primary jet there are secondary jets that can be central or forward with lower $p_T$ requirements.  
 To avoid events falling between searches,  a classification scheme that is relatively insensitive to the exact values used in the division of events is necessary.
 This is accomplished by selecting a common definition of a secondary jet and then classifying events by the number of secondary jets.    The $p_T$ selection of the jets is as shown in Table~\ref{Tab: Classification}.

\begin{table}
\centering
\begin{tabular}{|c||c|c|c|c|}
\hline
&\multicolumn{4}{c|}{Selection Criteria for Different Jet Multiplicity Searches}\\
\hline
      & $1j+\MET$& $2j+\MET$& $3j+\MET$            & $4^+j+\MET$\\
\hline\hline
$j_1$ &  $\ge 400\GeV$&  $\ge 100\GeV$&  $\geq 100\GeV$        &  $\geq 100\GeV$ \\
$j_2$ &  $< 50\GeV$&  $\ge 50\GeV$&  $\ge  50\GeV$       &  $\geq 50\GeV$ \\
$j_3$ &  $< 50\GeV$&  $<50\GeV$&  $\ge 50\GeV$        &  $\geq 50 \GeV$ \\
$j_4$  &  $< 50\GeV$&  $<50\GeV$&  $<50\GeV$       &  $\geq 50 \GeV$ \\
\hline
\end{tabular}
\caption{The selection criteria for the different jets plus missing energy samples. The abundance of ISR and FSR means that only the $4^{+}j+\MET$ channel is necessary to maximize sensitivity.
 Jet $p_T$ cuts in GeV for the different exclusive search channels. The following cuts are applied: $|\eta_j| <2.5$, $p_{T,\ell}<20$ GeV, $\Delta\phi(\MET, j_i)>0.2$ rad. ($i=1,2,3$).}
\label{Tab: Classification}
\end{table}

There is a trade-off between having a higher versus lower requirement for jet $p_T$s.  Higher jet $p_T$s mean that the jets are more likely to come from final state decay products and less  likely to be produced in parton showering and radiation.  Having a tighter requirement on jet $p_T$s will reduce the SM backgrounds significantly. However, new physics may not have a widely spaced spectrum and the resulting final state jets may be soft.  Therefore, by raising the requirements on  jet $p_T$s, there is a loss of sensitivity to degenerate spectra.    
At the same time, the search strategy should  pick out spectacularly hard jets from a sea of soft jets.  In Sec. \ref{Sec: Observables}, other event variables, namely $H_T$ or $M_\eff$, are used to separate very energetic events.    

An alternative strategy would be to incorporate multiple searches using different requirements on jet $p_T$s in different searches.   While this approach will increase reach, it does so at the expense of rapidly increasing the number of searches necessary to cover the full range of possible signals. This alternative was studied in this article; ultimately, there is little gain in the discovery potential for this more complicated search.  Therefore,  keeping the jet $p_T$ requirements fixed and relatively modest is a good compromise and using $H_T$ or $M_\eff$ as a way of effectively increasing the $p_T$ requirements on the jets.    The value of the lower multiplicity searches is diminished by having  modest $p_T$ requirements.  The reduced value of lower multiplicity searches has a side benefit:  a smaller set of searches is necessary to cover the full range of new physics possibilities.  Thus, a multijet search will be most effective even if the decay products of the parent particle are effectively invisible and all the jets must be generated through radiation.

\subsection{Observables}
\label{Sec: Observables}

The next step in the search strategy is to use kinematic or event shape variables to discriminate signal from background.  Minimizing the number of searches performed and the number of event shape variables used is desirable  for several reasons.  

First, the full scope of  beyond the Standard Model signals  frequently populates most of the spectrum of these kinematic variables.    It is necessary to examine multiple cut values for each kinematic variable used in order to not eliminate signal.  For instance, \cite{GeneralizedJetsMET} showed that  the high $H_T$  cuts used in some jets plus missing energy searches at the Tevatron could eliminate potentially discoverable signals.  
Using more variables results in an increase in the number of searches necessary to discover generic beyond the Standard Model signals.  An additional reason to avoid too many kinematic variables in a search is that higher order corrections to the signal and background may alter the expected sensitivity, leading to greater theoretical uncertainty in the searches.  The more kinematic variables used, the greater this problem is exacerbated.

Three event shape variables are useful at separating signal from background and are studied below.  The squark and gluino benchmark models are used to test  the efficacy of cuts on different event shape variables that enhance visibility of new physics over background.  A series of three searches are necessary to have broad sensitivity to beyond the Standard Model physics.

 Intuitively, $\MET$ is an important event shape variable to use in discriminating signal from background. Moreover, additional resolving power can frequently be attained by using a second kinematic variable such as  $M_{\text{eff}}$ or $H_T$.
In addition to the tried-and-true kinematic variables, the sensitivity in multijet and $\MET$ searches is explored by using the event shape variable, $\arts$, first introduced in \cite{Randall:2008rw} as an alternative to using a fixed $\MET$ cut for removing the QCD backgrounds in dijet and missing energy searches. 
$\arts$ was initially defined as
\begin{eqnarray}
\label{Eq: AlphaRTS}
\arts &=& \frac{p_{T{j_2}}}{m_{j_{1}j_{2}}}
\simeq \sqrt{\frac{p_{Tj_2}}{2 p_{Tj_1}(1 - \cos{\theta_{j1j2}})}}
\end{eqnarray}
where $p_{T_{j_{1,2}}}$ are the transverse momentum of the leading two  jets in the event,  $m_{j_{1}j_{2}}$ is the invariant mass between the leading and subleading jets, and $\theta_{j1j2}$ is the angle between them. 

In  QCD dijet events, where the jets are predominantly back-to-back, Eq.~\ref{Eq: AlphaRTS} gives $\arts$=0.5.  Fluctuations of a jet's energy reduce $\arts$ and typical QCD events have $\arts \le 0.5$.   A SUSY signal can extend beyond $\arts$ of 0.5 because $\MET$ is not being created due to fluctuations of the jet energy  \cite{Randall:2008rw}.   This suggested that a cut of $\arts > 0.5$ could be used in lieu of a $\MET$ cut in dijet events.    However, as discussed in Sec. \ref{Sec: Search Channels}, that radiation typically produces several additional jets in addition to the signal and reduces the efficacy of the dijet search.

\begin{figure}[htb]
\begin{center}
 \includegraphics[width=6.in]{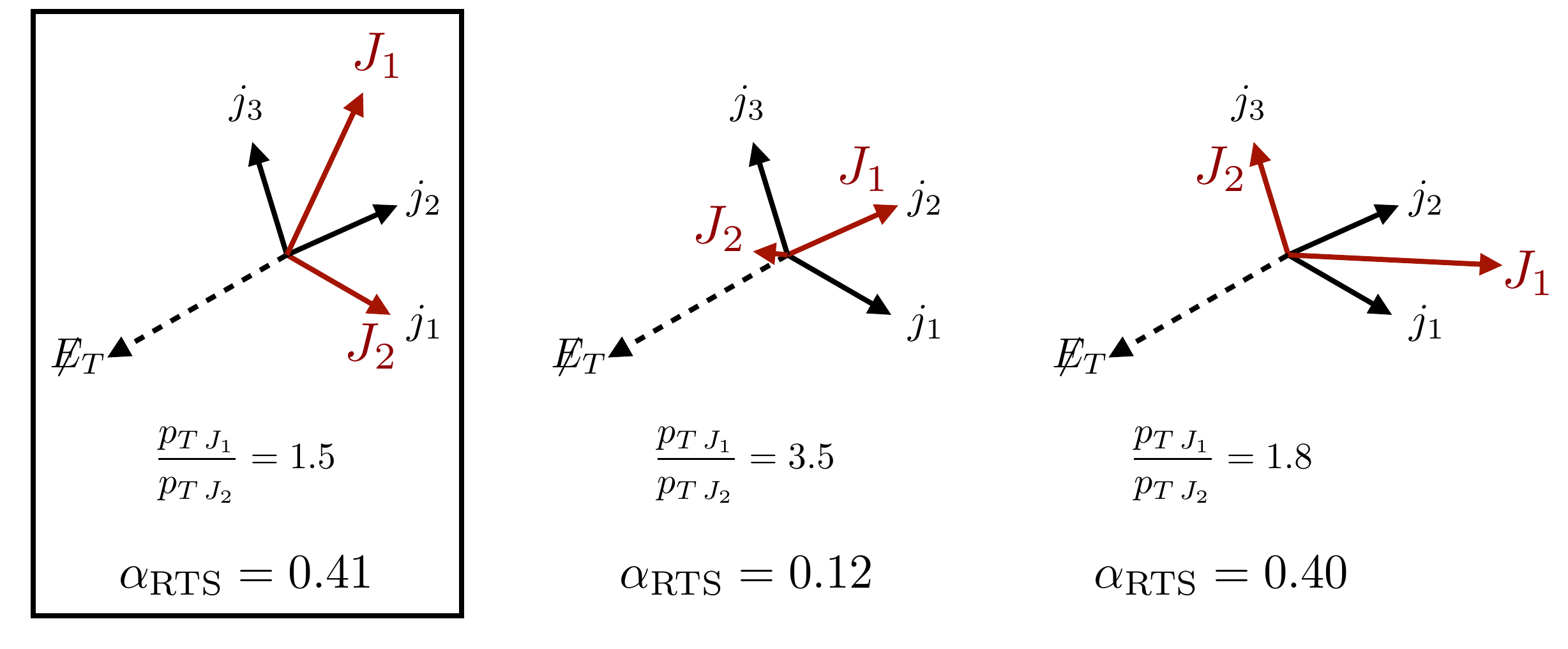}
 \caption{A sample three jet ($j_1, j_2, j_3,$) plus $\MET$ event with the $p_T$ ratios and $\arts$ computed for each of the three  superjet groupings.    The grouping that minimizes the $p_T$ imbalance of  the two superjets is chosen and $\arts$ is computed from these two superjets. In this case, the first grouping has the lowest momentum imbalance.}
   \label{Fig: SuperJets}
   \end{center}
 \end{figure}

Initial studies for dijet and missing energy searches at the LHC indicate that $\arts$ is an effective event shape variable for dijets \cite{CMS_note, Baer:2009dn}; however, exclusive dijet searches are never the most effective discovery channel for the classes of signals considered in this article even using $\arts$.   $\arts$ can be generalized \cite{Alpha_Generalized} to events with multijets by assigning each of the jets to one of two groups of jets, or superjets ($J_1$ and $J_2$), so that the ratio $p_{T\,J_1}/p_{T\, J_2}$ is minimized.  $\arts$ is then calculated with Eq.~\ref{Eq: AlphaRTS} using the 4-momentum of superjets $J_1$ and $J_2$ rather than the leading jets. The grouping of jets into superjets is illustrated in Fig.~\ref{Fig: SuperJets}.

As an illustrative example, consider a 700 GeV gluino directly decaying to a 350 GeV LSP. The left panel of Fig.~\ref{AlphaQCD} shows a $\arts$ distribution for a simulated QCD sample and the electroweak SM backgrounds with no $\MET$ cut (top panel); its shape retains the feature $\arts\geq 0.50$ as an alternative to a $\MET$ cut in dijets and $\MET$ searches.  However, many new physics scenarios do not have a significant tail beyond 
$\arts=0.50$.    Using  $\arts$ in lieu of a $\MET$ requirement was not sufficient; however, using $\arts$ and a moderate $\MET$ could improve the searches over a $\MET$ cut alone.     The right panel of Fig.~\ref{AlphaQCD} shows the $\arts$ distributions with a $\MET$ cut of 400 GeV.  The signal begins to deviate from backgrounds at moderate $\arts \simeq 0.20$ and by cutting  at $\arts \geq 0.20$, the signal is discoverable with $\LL=1 \ifb$.

\begin{figure}[htb]
\begin{center}
 \includegraphics[width=3.in]{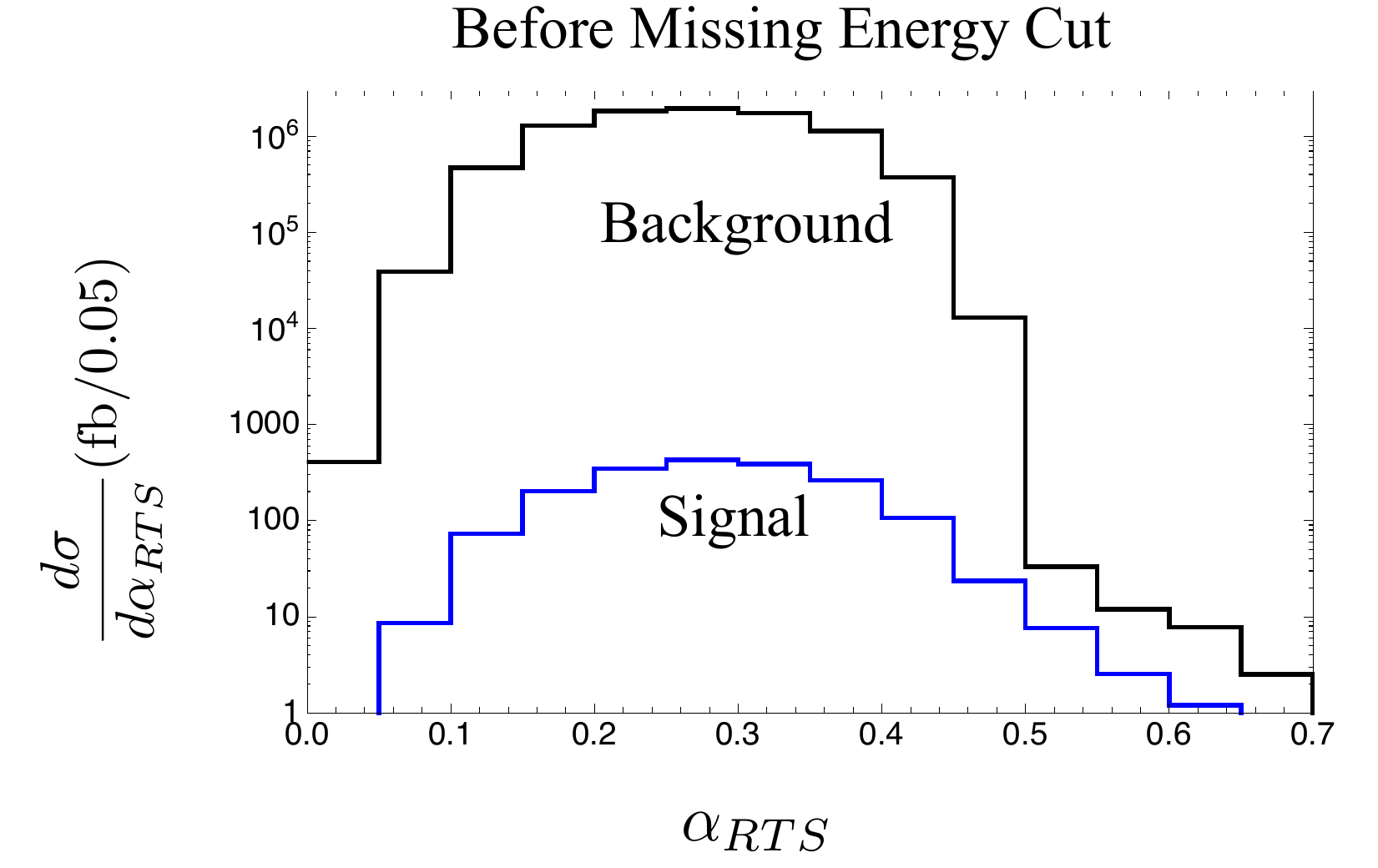}
  \includegraphics[width=3.in]{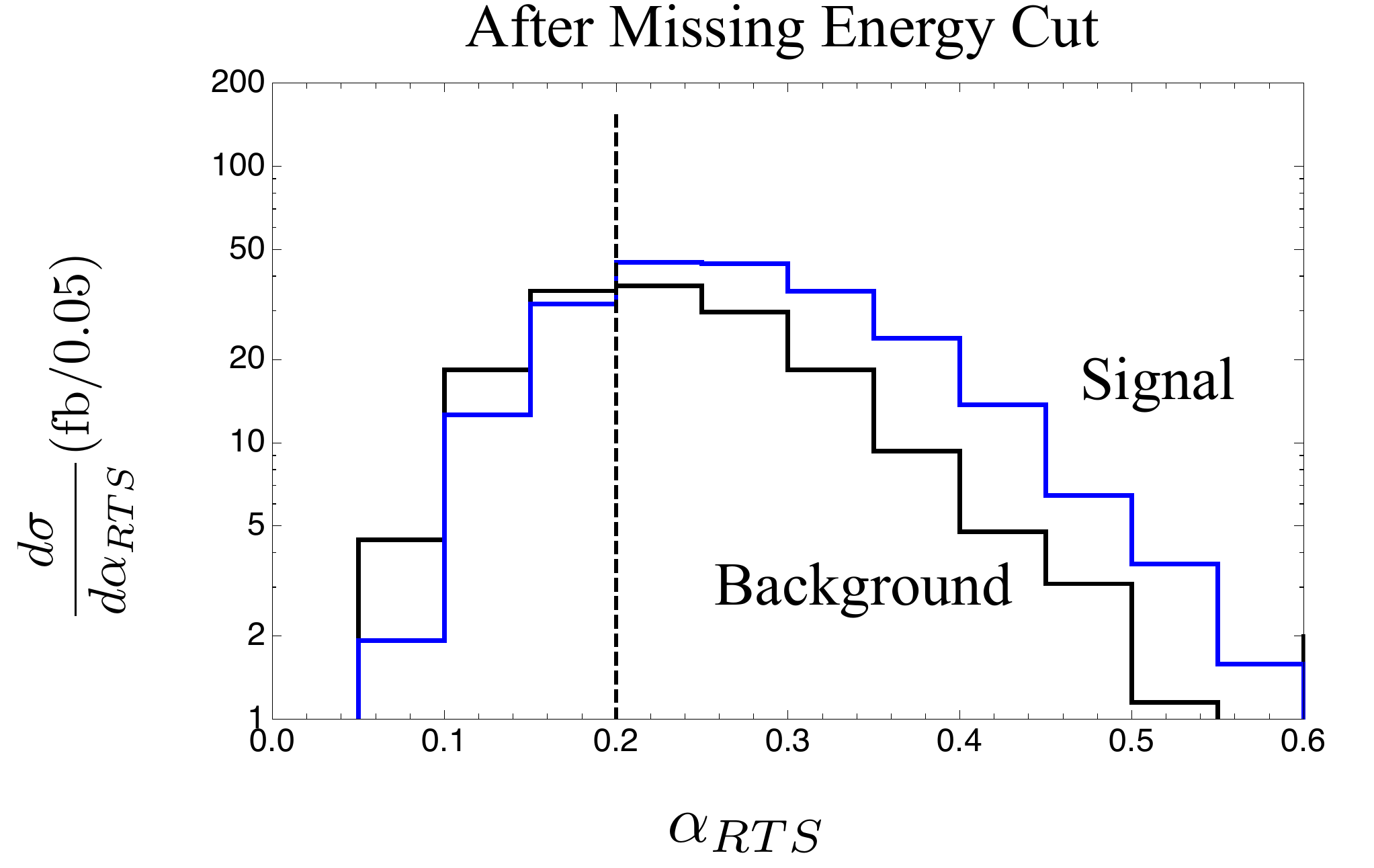}
 \caption{$\arts$ distributions of combined SM backgrounds (black) and a  signal spectrum with a 700 GeV gluino decaying to a 350 GeV LSP (blue).  The left panels shows the $\arts$ distribution with no $\MET$ requirement.   The right panel shows the distributions after a  $\MET \ge 400 \GeV$ is imposed. Using $\arts> 0.20$ in conjunction with $\MET$ in a set of searches allows for discovery with the criteria in Sec. 4.}
   \label{AlphaQCD}
   \end{center}
 \end{figure}

Ultimately three sets of searches are needed to separate signal from background. For the gluino pair-production case, additional coverage along the degeneracy region, $m_{\chi^0}\lsim m_{\tilde{g}}$ is gained by considering a set of multiple cuts on $H_T$ while applying a high $\MET\geq 600 \GeV$ cut and no $\arts$ cut $(\alpha_{\text{RTS}}\geq 0)$. The ``High $\MET$  Searches'' also maximizes the reach of a search for pair-produced squarks when $m_{\tilde{q}}\gsim m_{\chi^{0}}$.  A second set of searches, the ``$\arts$ Searches,'' in $H_T$ keeping $\MET\geq 400 \GeV$ but requiring $\arts\geq 0.2$ picks up those signal points of high mass gluinos with $2 m_{\chi^0} \lsim m_{\tilde{g}}$, and the large squark mass low LSP mass region. An additional set of searches, the ``Base Searches,'' in $H_T$ while keeping a medium $\MET\geq400\GeV$ cut and no $\arts$ cut, is necessary to achieve discovery reach for those theories where the energy goes into visible states. For example, consider a theory where gluinos are pair-produced, but cascade decay via a Wino to the LSP and jets. The selection criteria for each of the three sets of searches is illustrated in Table~\ref{Tab: Different Searches}. Lowering the  $\MET$ requirement does not enhance the discovery potential due to the steeply rising background and by $\MET \ge 600 \GeV$, the backgrounds are down to $\OO(10\fb)$. The SM backgrounds $\MET$ spectrum falls faster than a typical signal point, thus lower cuts on $\MET$ only decrease $S/B$.  These searches are not exclusive and cannot be combined together; however, each set is particularly suited for enhancing coverage in a  region in the $(m_{\tilde{g}}, m_{\chi^0})$ mass parameter space. These strategies are applied to estimate the discovery sensitivity at the LHC at $\sqrt{s}=14 \TeV$ for pair-produced gluinos that either cascade or directly decay to the LSP and two jets, and to pair-produced squarks that decay directly to the LSP and a jet.

\begin{table}[ht]
\begin{center}
\begin{tabular}{|c||c|c||c|}
\hline
 & $\MET$ & $\arts$&$H_T$\\
\hline \hline
Base& $\ge 400 \GeV$& $\ge 0.0$& $ \ge$ 600 GeV, 900 GeV, 1200 GeV, 1500 GeV\\
\hline
High $\MET$& $\ge 600 \GeV$& $\ge 0.0$& $ \ge$ 600 GeV, 900 GeV, 1200 GeV, 1500 GeV\\
\hline
$\arts$& $\ge 400 \GeV$& $\ge 0.2$& $ \ge$ 600 GeV, 900 GeV, 1200 GeV, 1500 GeV\\
\hline
\end{tabular}
\caption{ \label{Tab: Different Searches}
The different multijet plus missing energy searches used.  The left table shows the division of searches
into the ``High $\MET$'' search, and ``$\arts$'' search.
The right table shows the different supplementary  $H_T$ cuts necessary to gain
full coverage.  The larger $H_T$ cuts are effective at looking for highly non-degenerate
spectra.  
}
\end{center}
\end{table}

\section{Signal and Background Calculation}
\label{Sec: Generation}

This section describes the calculation of signal and background.  This article uses leading order Monte Carlo calculations
for processes up to $2 \rightarrow 4$.  The cross sections are scaled to the NLO inclusive cross section.

The dominant backgrounds to $\MET$ + jets searches are $t\bar{t}+nj$, $W^{\pm}+nj$, $Z^0+nj$, and QCD. Subdominant contributions from $W^{\pm}W^{\pm}$, $Z^{0}W^{\pm}$, single top, and $Z^{0}Z^{0}$
are not considered in this article.

Additional jets are calculated through two distinct approximation techniques in the Monte Carlo event generation.
The first method uses Feynman diagram techniques with larger multiplicity final states. 
This approximation procedure gets the wide-angle, hard emission correct, but
is computationally expensive and does not resum larger logarithms of kinematic variables.
The parton shower is the second approximation procedure used to produce additional final state
jets.  Parton showers are appropriate in the collinear approximation and resum larger logarithms, but
underestimate the amount of hard radiation.  Both approximation schemes are used in Monte Carlo
generators and matching final state jets to matrix element partons is necessary to avoid double counting a process represented in both methods. 

The results of this study will rely on the $4^{+}j + \MET$ channel. Ideally, one would want to generate up to four jets at matrix element level, but this is a computationally impractical task. Jets beyond those coming from the matrix element can be reasonably well approximated by the parton shower. For example, in \cite{Alwall:2007fs}, $W^{+}+jets$ were matched up to four jets. In this article, this background is matched up to three jets. The ratio of the inclusive $W^{+}+4j$ rate to that of the inclusive $W^{+}+3j$ rate from the samples generated for this study differ from the ratio from \cite{Alwall:2007fs} by $\mathcal{O}(15\%)$.

This article uses a MLM-based \cite{PSME} $k_\perp$ shower matching scheme \cite{SUSYPSME}, where partons are clustered using the $k_{\perp}$-algorithm.
The $k_\perp$ jet measure is defined as
\begin{equation}
d^2(i,j) = \Delta R^2_{ij} \min(p_{T\, i}^2. p_{T\, j}^2),
\end{equation}
 where
\begin{equation}
\Delta R^2_{ij} = 2(\cosh\Delta\eta_{ij} - \cos\Delta\phi_{ij}).
\end{equation}
At the matrix element level, partons are required to be separated by a minimum cut-off $Q^{\text{ME}}$ \cite{KT}. After the parton shower and hadronization, the final state hadrons cluster into jets.  Matching consists of identifying a jet with a parton. There are two instances to consider. First, for lower jet multiplicity samples, a jet is considered matched when it is within $Q^{\text{PS}}$ of a parton, namely $d(\text{parton}, j)<Q^{\text{PS}}$. If the jet cannot be matched, the event is discarded. Second, in the sample with the highest jet multiplicity (2 jets for $t\bar{t}$ and signal, 3 jets for $W^{\pm}$ and $Z^{0}$), additional unmatched jets are allowed provided that they are softer than the softest parton.
The $k_\perp$ shower scheme sets $Q^{\text{ME}}=Q^{\text{PS}}$ and 
in order to generate higher $p_T$ jets,
it is necessary to set the minimum $p_{T}$ of the jets at parton level equal to $Q^{\text{ME}}$.

For the studies in this article, events were generated using \texttt{Madgraph/MadEvent 4.4.0 (4.4.2)}\cite{Alwall:2007st} for each of the following background processes:
\begin{eqnarray}
\nonumber
pp  \rightarrow W^{\pm} + nj  &\quad& 1\le n\le 3\\
\nonumber
pp  \rightarrow Z^{0} + nj &\quad & 1\le n \le 3 \\
pp  \rightarrow t\bar{t} + nj &\quad & 0\le n \le 2 .
\end{eqnarray}
In order to have sufficient statistics at high missing energy, each of these backgrounds was calculated with
different requirements of the $p_T$ of the heavy particle.   For instance, the $Z^0+ nj$ was calculated in four samples
satisfying 
\begin{eqnarray}
\nonumber
p_{T\, Z^0} < 100 \GeV,&\qquad&
100\GeV \le p_{T\, Z^0} < 200 \GeV,\\
200 \GeV\le p_{T\, Z^0} < 400 \GeV,&\qquad&
400 \GeV\le p_{T\, Z^0}.
\end{eqnarray}
For each squark and gluino signal process considered, 100K events were produced:
\begin{eqnarray}
\nonumber
pp\rightarrow \widetilde{g} \widetilde{g} + nj \quad 0\le n \le 2\\
pp\rightarrow \widetilde{q} \widetilde{q}^{\dagger} + nj \quad 0\le n \le 2
\end{eqnarray}
\texttt{Pythia 6.4} is used to include initial and final state radiation beyond the radiation explicitly computed with \texttt{Madgraph/MadEvent}, and for the decays of all unstable particles  \cite{Sjostrand:2006za}. Finally, the ATLAS version of \texttt{PGS 4} \cite{PGS}, with $\Delta R=0.7$, was used as a detector simulator.
Appropriate $K$ factors were included to ensure that the leading order in \texttt{Madgraph/MadEvent}   production cross sections result in the correct total inclusive production cross section.
This article computed the $K$ factors by comparing LO cross sections to theoretical predictions of the NLO cross section.  As a cross check, after running the rescaled signal through {\tt PGS}, different kinematic distributions that have been published by ATLAS were checked \cite{Akimoto:2009zz}.
The calculations matched published distributions  in \cite{Akimoto:2009zz} up to residual $K$ factors that can be attributed to inaccurately modeled detector efficiencies such as lepton veto efficiencies in a multijet environment.

As illustrated in Fig.~\ref{matching}, the generation of higher multiplicity jets from the parton shower underestimates the high $p_{T}$ and $\MET$ regions significantly, particularly in scenarios where $m_{\tilde{g}}\sim m_{\chi^0}$. In this region of parameter space, the jets coming from the decay of the gluino may be too soft to pass selection. Additionally, the two LSPs tend to line up back to back giving low $\MET$. However, this topology is enhanced by including ISR jets. Emission of a hard ISR jet boosts the gluino pair and typically creates a momentum imbalance that gives high values of $\MET$. 

\begin{figure}[h]
\begin{center}
 \includegraphics[width=4in]{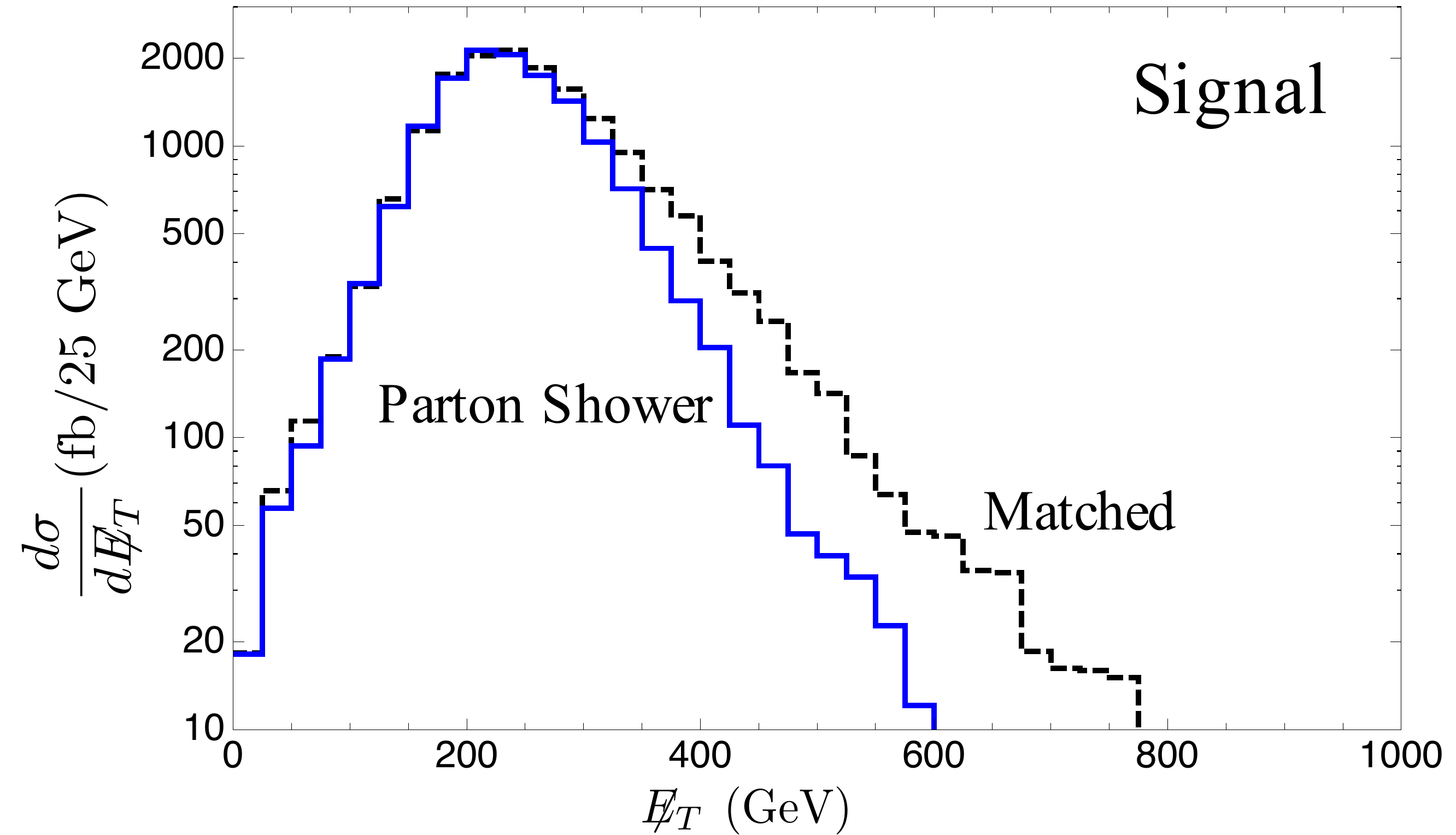}
  \caption{$\MET$ distributions for a matched (black) and unmatched (blue) 250 GeV gluino decaying to a 200 GeV LSP.}
   \label{matching}
   \end{center}
 \end{figure}

\subsection{Background Subprocesses}
\label{Sec: BG CS}

For both SM backgrounds and squark and gluino signals, {\tt CTEQ6L1} parton distribution functions \cite{Pumplin:2002vw} are used along with variable renormalization and factorization scales. This variable choice sets the scales to the central value of the transverse mass, $m_T$, of the event. At generation level at least one jet produced in association with $X=t\bar{t}, W^{\pm}, Z^{0}, \text{signal}$, must have $p_T > 100$ GeV for each of the processes $pp\rightarrow X + nj$ for $n\geq1$.

The backgrounds calculated in Sec. \ref{Sec: BG CS} are combined into a final estimate for the backgrounds
in the three series of searches.  The results are listed below in Table \ref{Grids4}.   This article uses
a benchmark luminosity for the searches of $1 \ifb$ and therefore these searches should have relatively
modest backgrounds.   In Sec. \ref{Sec: Results}, the sensitivity studies are performed and a systematic
error of 30\% is taken on all these cross sections.

\subsubsection*{$t\bar{t}+nj$}

$t\bar{t}+nj$ is the dominant background to three- and four-jets and $\MET$ searches at moderate values of missing energy (see Fig.~\ref{multijet}).  At matrix element level, the minimum transverse momentum of the jets is set to $p_{T_j}\geq100$ GeV $=$ $Q^{\text{ME}}$. Fig.~\ref{multijet} shows $\MET$ distributions for each SM background in the multijet channel. The LO cross section for $t\bar{t}+nj$ obtained is $\sigma = 483$ pb. A $K$ factor of 1.5 is taken from \cite{Kidonakis:2008mu}.  The $t\bar{t}$ $M_\eff$ distribution is compared to that of \cite{Akimoto:2009zz}, to obtain $K_{\eff, t\bar{t}}=1.62$. 

\begin{figure}[ht]
\begin{center}
 \includegraphics[width=4.5in]{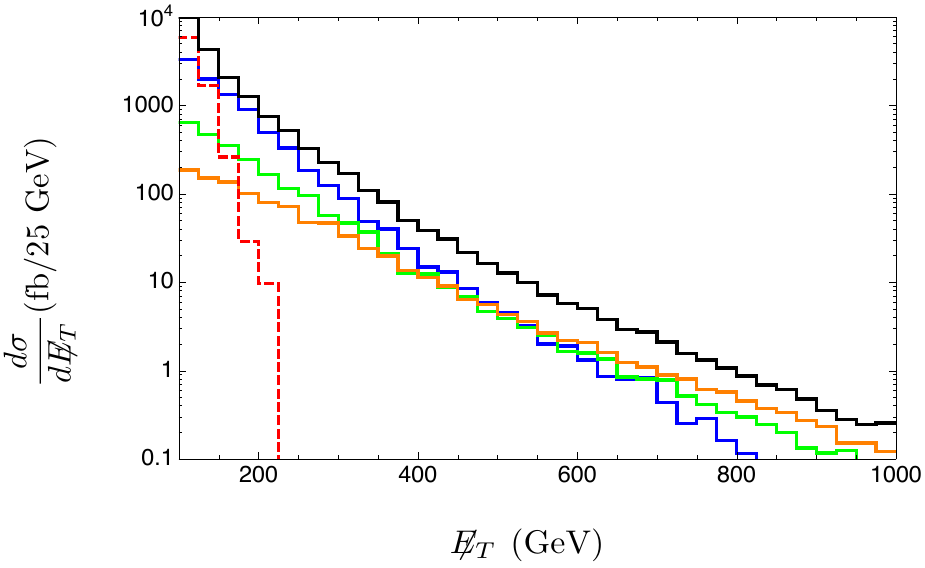}
  \caption{
    \label{multijet}
  $\MET$ distributions for $t\bar{t}$ (blue), $W^{\pm}$ (green),  $Z^0$ (orange), QCD (dashed red) and combined (black) in the multi-jet channel, after applying the selection criteria.}
  \end{center}
  \end{figure}

\subsubsection*{$W^\pm + nj$ and $Z^0+nj$}

$W^{\pm}$ and $Z^0$ backgrounds are the most relevant backgrounds for high values of missing energy (see Fig.~\ref{multijet}). At matrix element level, the minimum transverse momentum of the jets was set to $p_{T,jet}\geq 40$ GeV, and sets the matching scale to $Q^{\text{ME}}=Q^{\text{PS}}$.  Requiring that the $p_T$ of the  leading jet be at least 100 GeV does not conflict with the matching scheme if this cut is also placed at analysis level. To generate enough missing energy, only $W^\pm$ and $Z^0$ decay modes with final state neutrinos are calculated:
\begin{eqnarray}
W^{\pm}  \rightarrow \ell^{\pm} + \nu_{\ell} \qquad
Z^0  \rightarrow \nu \bar\nu  .
\end{eqnarray} 
Non-isolated muons are merged into the nearest jet and any objects {\tt PGS} identifies as hadronic taus are converted
into jets.

The LO cross sections are $460$ pb for $W^{+}$ $+$ $nj$, 303 pb for $W^{-}$ $+$ $nj$, and 166 pb for $Z^{0}$ $+$ $nj$.  $K$ factors are obtained from \cite{blackhatKfactors} for $W^{\pm}$ and found to be 1.38 for $W^{+}$ and 1.35 for $W^{-}$. These results agree with those obtained by comparison with \cite{WZkfactors}, and the effective $K$ factor computed by matching the $M_{\eff}$ to that from \cite{Akimoto:2009zz}. 

$Z^{0}+nj$ $K$ factors are not available in the literature for sufficiently high multiplcities. In this article an effective $K$ factor of 1.24 for $Z^{0}+nj$ is obtained by comparing the $M_\eff$ distribution to that in \cite{Akimoto:2009zz}. 

\subsubsection*{QCD}

The QCD background gives dominant low missing energy contributions.  The QCD background arises from a myriad of sources such as semileptonic decays of heavy quarks and  mismeasurements or instrumental effects at detector level. 
Simulating QCD is challenging given the uncertainties from detector effects.   
 A requirement on $\MET\ge 400 \GeV$ is placed to remove the QCD backgrounds in the searches in this article.
 This cut is significantly stronger than many proposed searches that are frequently $\MET \ge 200 \GeV$.  The tighter requirement makes QCD sources of missing energy unimportant for the searches in this article. 
The QCD contributions for this article are produced in \texttt{Madgraph/MadEvent 4.4.3}; for the following parton level process:
\begin{eqnarray}
pp\rightarrow jjjj
\end{eqnarray}
Each jet is required to have $p_T> 50 \GeV$, and at least one of them must have $p_T > 100 \GeV$. Additional jets come from showering and radiation done in \texttt{Pythia}. To generate enough statistics, the parton level events were divided into independent samples described by  $p_T$ cuts on the leading and sub-leading jets. 
The LO cross section obtained was $\sigma_{\text{QCD}} = 72 \text{ nb}$.


The dominant contribution to  $\MET$ from QCD arises from two sources: fluctuations of jet energies and neutrinos from heavy flavor decays.
Jet energy fluctuations dominate low $\MET$ and can compete with $t\bar{t}$, but is steeply falling.
The QCD sample generated from this article used \texttt{PGS} to model the fluctuations jet energy to generate $\MET$.   
Heavy flavor decays with real neutrinos dominate the tail of the QCD contribution to the $\MET$ distribution. 
No special attempt to quantify the heavy flavor contributions to missing energy; however, this tail  takes over at $\MET > 150 \GeV$, where QCD is subdominant to the high-$p_T$ sources of MET.
All the searches in this article require $\MET\geq 400 \GeV$ where the QCD/heavy flavor contributions to $\MET$ are well under control.

\section{Expected Sensitivity}
\label{Sec: Results}

 \begin{table}[h]
\centering
\begin{tabular}{|c||c|c|c|}
\hline
\multicolumn{4}{|c|}{Standard Model Cross Section}\\
\hline
$H_T$&Base&High $\MET$ & $\arts$\\
\hline\hline
$\ge 600\GeV$ & $155\fb$ & $25\fb $&$63\fb$\\ \hline
$\ge 900\GeV$ & $92\fb$& $21\fb$&$37\fb$\\ \hline
$\ge 1200 \GeV$ & $40\fb$&$ 12.6\fb$&$17\fb$\\ \hline
$\ge 1500\GeV$ & $16\fb$& $6.2\fb$&$8.0\fb$\\\hline
\end{tabular}
\caption{SM backgrounds cross sections (fb) in the multijet $+$ $\MET$ channel performing the ``High $\MET$" and the ``$\arts$ searches". }
\label{Grids4}
\end{table}

This section demonstrates the efficacy of the search strategy detailed in Table~\ref{Tab: Different Searches}. The abundance of ISR and FSR jets means that one need only consider the $4^{+}j+\MET$ channel. The sensitivity to several minimal particle contents are shown in Sec. \ref{Sec: Gluinos} to Sec. \ref{Sec: Squarks}.
This article performs counting experiments and discovery requires
\begin{eqnarray}
S \ge 5 \sigma_{\text{tot}} (B)
\end{eqnarray}
where $S$ is the  number signal events, $B$ is the number of background events, and $\sigma_{\text{tot}}(B)$ is
the combined statistical and systematic uncertainty in the background.

The three classes of searches introduced in Table~\ref{Tab: Different Searches} are used to place discovery reach bounds for different sample theories. The first step is to calculate the cross section of the Standard Model backgrounds for each of the search channels.  
Table~\ref{Grids4} shows the Standard Model predictions for the $3\times 4$  searches.  

\begin{table}[ht]
\centering
\begin{tabular}{|c||c|c|c|}
\hline
\multicolumn{4}{|c|}{Minimum Cross Section for 5 $\sigma$ Discovery}\\
\hline
$H_T$&Base&High $\MET$ & $\arts$\\
\hline\hline
$\ge 600\GeV$ & $241\fb$ & $45\fb $&$103\fb$\\ \hline
$\ge 900\GeV$ & $146\fb$& $39\fb$&$63\fb$\\ \hline
$\ge 1200 \GeV$ & $68\fb$&$ 25.9\fb$&$33\fb$\\ \hline
$\ge 1500\GeV$ & $31\fb$& $15.5\fb$&$18.5\fb$\\\hline
\end{tabular}
\caption{The minimum cross section for a $5\sigma$ discovery in the $4^{+}j+\MET$ channel when including Poisson fluctuations and a 30\% systematic uncertainty.}
\label{Tab: Discovery CS}
\end{table}

The uncertainty in any measurement is given by
 \begin{eqnarray}
\sigma_{\text{tot}}(B)= \sqrt{\sigma_{\text{Pois}}(B)^{2}+ (\epsilon_{\text{Syst}} B)^{2}},
 \end{eqnarray}
 where $B$ is the number of background events for a given measurement. The statistical error  is Poisson with a variance, $\sigma_{\text{Pois}}(B)$.  $\sigma_{\text{Pois}}(B)$ approaches $\sqrt{B}$ for large backgrounds.
 In addition to statistical fluctuations of backgrounds,  there are systematic uncertainties that don't proportionally reduce with increased luminosity.
 $\epsilon_{\text{Syst}}$ is the systematic uncertainty in the prediction and measurement and will ultimately limit the search reach.   This systematic background represents both the theoretical uncertainty in the backgrounds and the experimental systematics. In the early running of the LHC a value of $\epsilon_{\text{Syst}}=30\%$ is a plausible estimate of the systematic uncertainty and is used throughout this article \cite{Polesello}.   In order to discovery a signal with these requirements, a signal needs to have a cross section $\sigma(S) \ge 5\sigma_{\text{tot}}(B)$.  These cross sections are listed in Table~\ref{Tab: Discovery CS}.

\begin{table}[ht]
\centering
\begin{tabular}{|c||c|c|c|}
\hline
$H_T$&Base&High $\MET$ & $\arts$\\
\hline\hline
$ \ge 600\GeV$ & $211\fb$ & $35\fb$ &$121\fb$\\ \hline
$\ge 900\GeV$& $137\fb$& $30\fb$&$72\fb$\\ \hline
$\ge 1200\GeV$& $62\fb$& $19\fb$&$31\fb$\\\hline
$\ge 1500\GeV$& $26\fb$& $10\fb$&$13\fb$\\\hline
\end{tabular}
\caption{
Cross sections in each search channel for a $(m_{\tilde{g}}, m_{\chi^0})=(800\GeV, 300\GeV)$ sample spectrum. The $H_T = 600\GeV, 900\GeV$ $\arts$ searches have sufficiently large cross sections to be discovered.}
\label{signal1grid}
\end{table}

 Up to this point, the procedure is entirely model-independent. The next step is to take the model of choice and compute the number of events that it predicts for each of the searches described in Table~\ref{Tab: Different Searches}.
As an illustration of how a search for a signal point should proceed, Table~\ref{signal1grid} displays the multijet cross sections for a 800 GeV gluino decaying to a 300 GeV LSP. The greatest significance is found by applying a cut $(H_T>600$ $\GeV, \arts>0.2)$, the $\arts$ search.

 The remaining portion of this article demonstrates  how these $3 \times 4$ searches listed in Table~\ref{Tab: Different Searches} can effectively cover the kinematic parameter space of both jet-rich signals like gluino pair production and jet-poor signals like squark pair production.    
 
\subsection{Gluino Sensitivity}
\label{Sec: Gluinos}

 The model-independent search strategy described above is applied to a search for pair-produced gluinos. Each gluino decays via:
 \begin{eqnarray}
 \tilde{g}\rightarrow q+\bar{q}+ \chi^0.
 \end{eqnarray}
 The masses of the gluino and the LSP, the most relevant experimental observables, are unconstrained. 
 The minimum transverse momentum of the ISR jets is set to $p_T\ge 100$ GeV at parton level. Each gluino can decay to any one of the lightest four standard model quarks and its antiquark. 
The gluinos are produced dominantly through QCD interactions.    The NLO cross sections were calculated in \texttt{Prospino 2.1} \cite{Prospino, OtherNLOSusy} and used to obtain $K$ factors for each gluino mass point generated. {\tt MadGraph} is used to generate additional hard radiation which is important for degenerate spectra.
The matching of matrix element radiation to parton shower jets is performed in an identical procedure to the SM $t\bar{t}$ backgrounds.

 The $(m_{\tilde{g}}, m_{\chi^0})$ mass parameter space is partitioned into three interesting regions, and although these subsets of mass parameter space are not exclusive, they serve to illustrate the efficacy of the sets of searches listed in Table~\ref{Tab: Different Searches}. First, the bulk region, where the mass splitting between parent and daughter particle is larger than the mass of the daughter particle and is a universal feature in mSUGRA models. In these benchmark spectra, the visible and missing energy of the jets coming from a SUSY decay are large enough for $H_{T, \text{signal}}$ to beat the SM backgrounds. Thus this region is accessible to multijet and $\MET$ searches through both the high $\MET$ and the $\arts$ searches.
  
The second region of interest is the degenerate spectra scenario where the mass splitting between the parent and daughter particle is less than the daughter particle mass.   Degenerate spectra can only be discovered because there are energetic ISR and FSR jets associated with their production.  Radiation from heavy objects is suppressed, and therefore most of the jets in heavier degenerate spectra come from ISR and are at larger rapidity.   The searches described in this article use a rapidity cut of $|\eta_j|\le 2.5$, which allows even forward jets to contribute to the signal.  

The degenerate gluino and LSP scenario has a slower falling distribution, as ISR and FSR jets populate this topology typically in regions of high $\eta$. In non-degenerate spectra, as in the 800 GeV gluino, the more central jets, both from FSR and decay, play a greater role. Having hard radiation jets recoil off of a heavy gluino particle boosts the gluino pair and creates a momentum imbalance which typically gives large $\MET$. This second scenario is discoverable through the high $\MET$ search. Additional discovery reach is possible due to the large gluino production cross section and degenerate spectra with masses up to 500 GeV, which are visible at the LHC with this set of searches.

The third subset of mass parameter of space to consider is when $m_{\chi^0}$ is roughly within an additive factor of $m_{\tilde{g}}$ (\emph{e.g.} $m_{\chi^0}\lsim 2 m_{\tilde{g}}$). This is a region where the reach can be extended performing an $\arts$ search. The low $\arts$ region, $\arts<0.2$,  vetoed in the search is largely populated by SM backgrounds.    Compressed spectra signals populate larger $\arts$ due to a greater energy mismatch in the jets from SM, pushing $\arts$ from Eq.~\ref{Eq: AlphaRTS} to lower values. With signals however, the energetic jets from radiation are leveled in energy with the energy from the decay jets. For instance, a 750 GeV gluino decaying to a 350 LSP gives roughly 200 GeV jets. This pushes $\arts$ in Eq.~\ref{Eq: AlphaRTS} to higher values than 0.2. The results are summarized in Fig. ~\ref{Fig: GluinoDiscoveryPlot}, which shows the discovery reach at the LHC with $\sqrt{s}=14 \TeV$ and an integrated luminosity of $1\ifb$ for pair-produced gluinos. 

\begin{figure}[htb]
\begin{center}
 \includegraphics[width=4.5in]{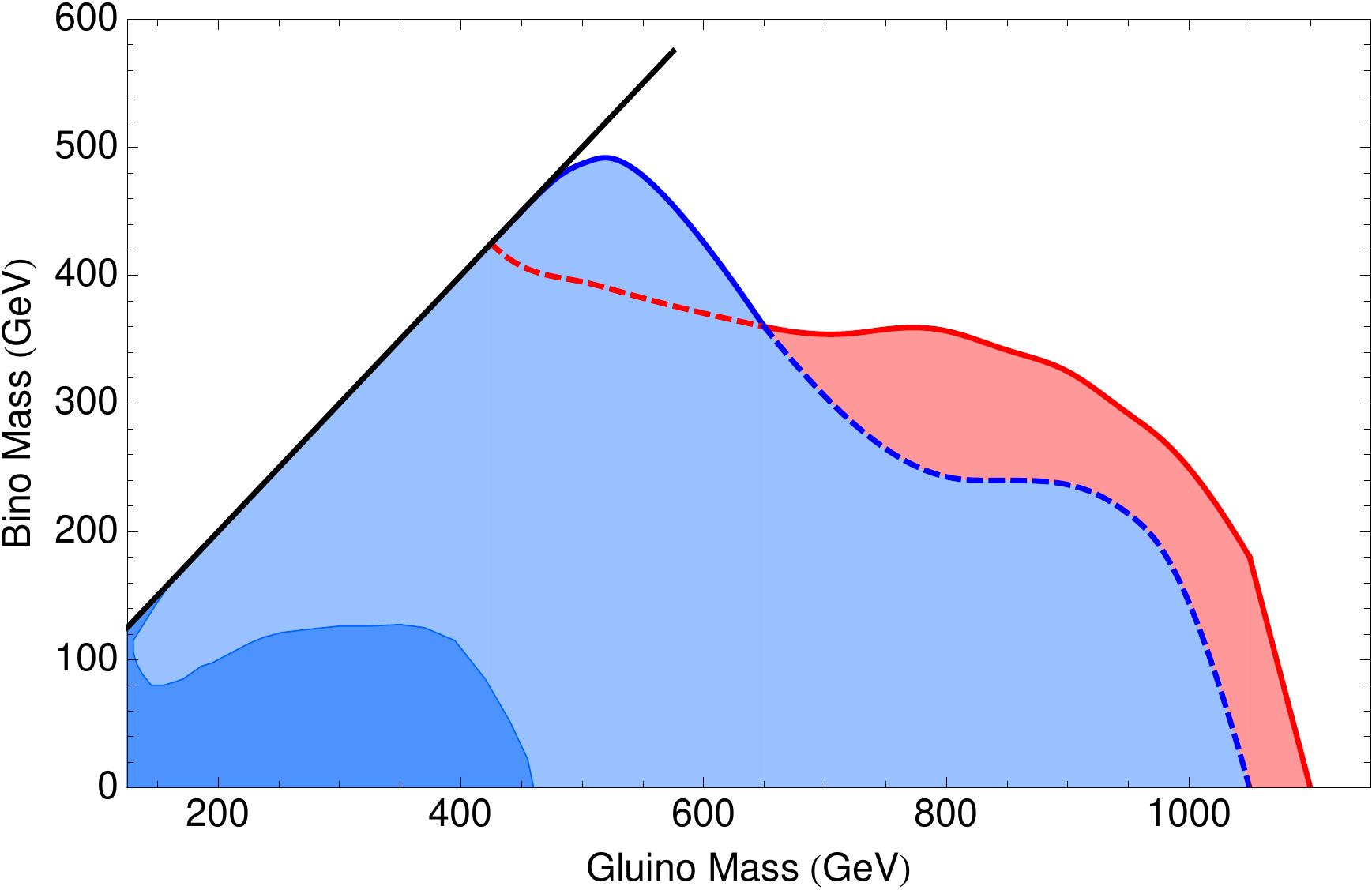}
  \caption{Discovery reach at the LHC for directly decaying gluino pair production at $\LL=1\ifb$. In light blue, the reach of a high $\MET$ set of searches, and the additional reach due to an  $\arts$ search (red).  Also shown is the $2\sigma$ exclusion at the Tevatron with 4 fb$^{-1}$ of data (dark blue) from \cite{ModelIndependentJetsMET}.}
   \label{Fig: GluinoDiscoveryPlot}
 \end{center}
 \end{figure}

 \subsection{Cascade Decays}
 The case studied so far assumes that each gluino decays directly to two jets and the LSP. It is important to consider an alternate scenario where the energy goes into intermediate states, \emph{i.e.} cascade decays. Here the gluinos are allowed to decay via:
  \begin{eqnarray}
 \tilde{g} \rightarrow q + q^{'}+ \tilde{W}\qquad 
 \widetilde{W} \rightarrow W^* + \chi^0.
 \end{eqnarray}
In the mass spectrum, the Wino is assumed to lie halfway between the gluino and LSP, namely $m_{\widetilde{W}} = 1/2 \left(m_{\tilde{g}} + m_{LSP}\right)$. This spacing in the mass spectrum was shown in \cite{GeneralizedJetsMET} to be the most pessimistic scenario at the Tevatron.  Cascade decays lower the $\MET$ of the event because the energy is distributed to more visible states. These theories require lowering the $\MET$ to have maximal sensitivity.  Fig.~\ref{Fig: CascadeDecayDiscoveryPlot} shows the coverage by the three sets of searches introduced. The reach of the high $\MET$ search turns off when there is enough phase space for the $W$ boson to go on-shell. 

\begin{figure}[htb]
\begin{center}
 \includegraphics[width=4.5in]{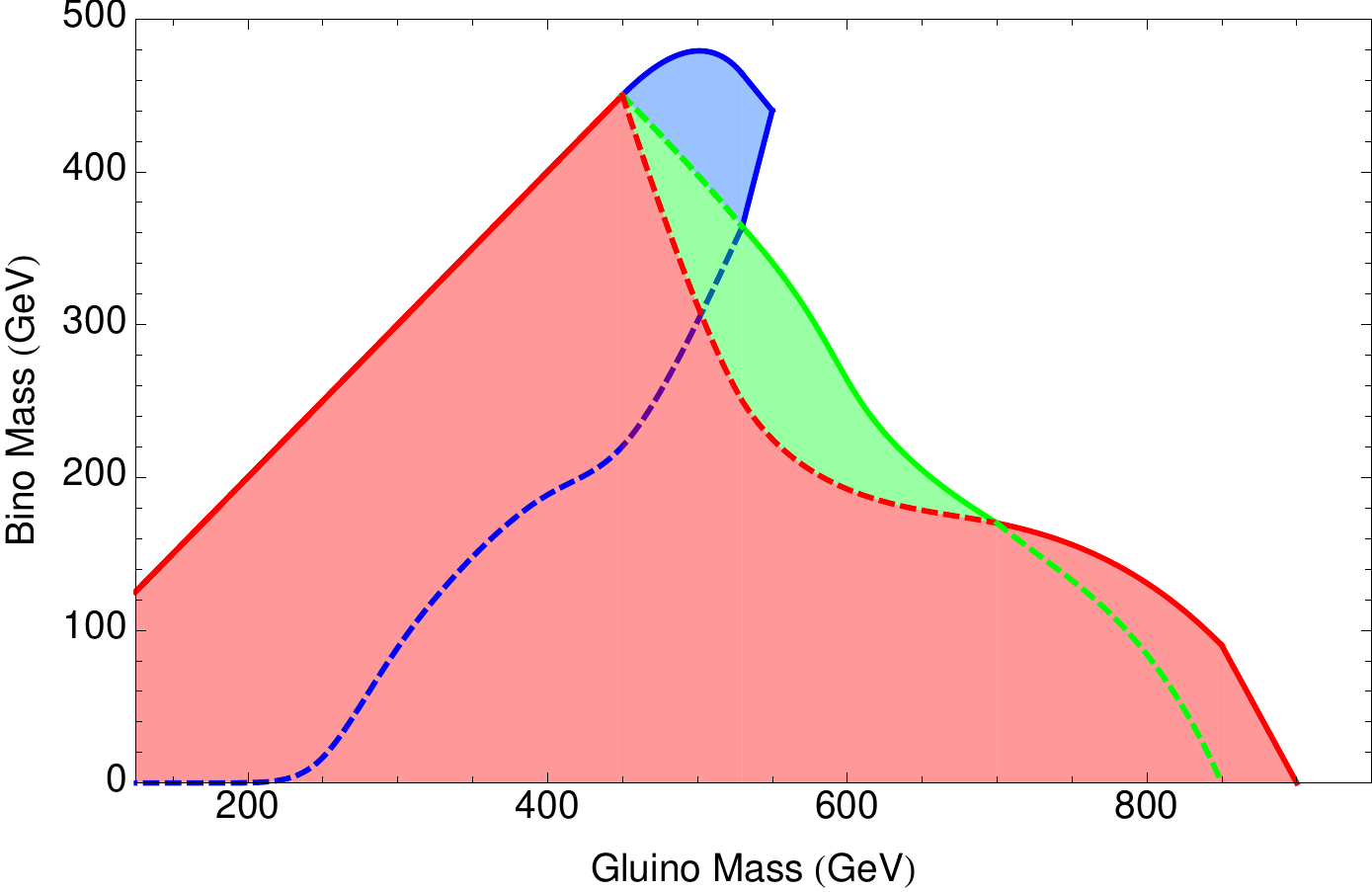}
  \caption{Discovery reach at the LHC for cascade-decaying gluino pair production at $\LL=1\ifb$. In light blue, the reach of a high $\MET\geq600\GeV$ set of searches, and the reach due to an  $\arts$ search (red). The reach of a Base Search $\MET\geq400\GeV$ is shown in green.}
   \label{Fig: CascadeDecayDiscoveryPlot}
 \end{center}
 \end{figure}

\subsection{Squark Sensitivity}
\label{Sec: Squarks}

 The third class of theories studied in this article include spectra where the squarks are significantly lighter than the gluinos.  
In this case, pair-produced squarks are the dominant mechanism for discovering new physics.  Squarks are scalars that are triplets of color, therefore, their production cross section and the associated radiation is smaller than gluinos.  As a way of studying the efficacy of the multijet search strategies proposed in this article, the squarks are forced to decay directly to the LSP:
 \begin{eqnarray}
 \tilde{q}\rightarrow q + \chi^0 .
 \end{eqnarray}
This topology indicates that a dijet plus missing energy search should be effective.  However, ISR and FSR jets play a crucial role in this case as well.  
The matching scale was set to $Q^{\text{ME}}=Q^{\text{PS}}=p_{T,j}\geq75$ GeV. A $K$-factor is used by calculating the NLO cross section in \texttt{Prospino 2.1} as in the case of gluinos \cite{Prospino, OtherNLOSusy}.
Parton shower/matrix element matching is performed upon the signal to generate additional radiation.  Due to the smaller color factors, there is significantly less overall radiation.  Nevertheless, the additional jets in the event and the moderate $p_T$ requirement on them are sufficient to supplement the two final state jets from squark pair production.  

A squark signal point and the background $\MET$ distributions for squarks are shown in Fig. \ref{Fig: Squark Comparison} for the multijet and dijet plus missing energy searches. While the overall  event rate of dijet and $\MET$ is larger than the multijet, the discovery potential is greater for the multijet search. 

 \begin{figure}[htb]
\begin{center}
 \includegraphics[width=3in]{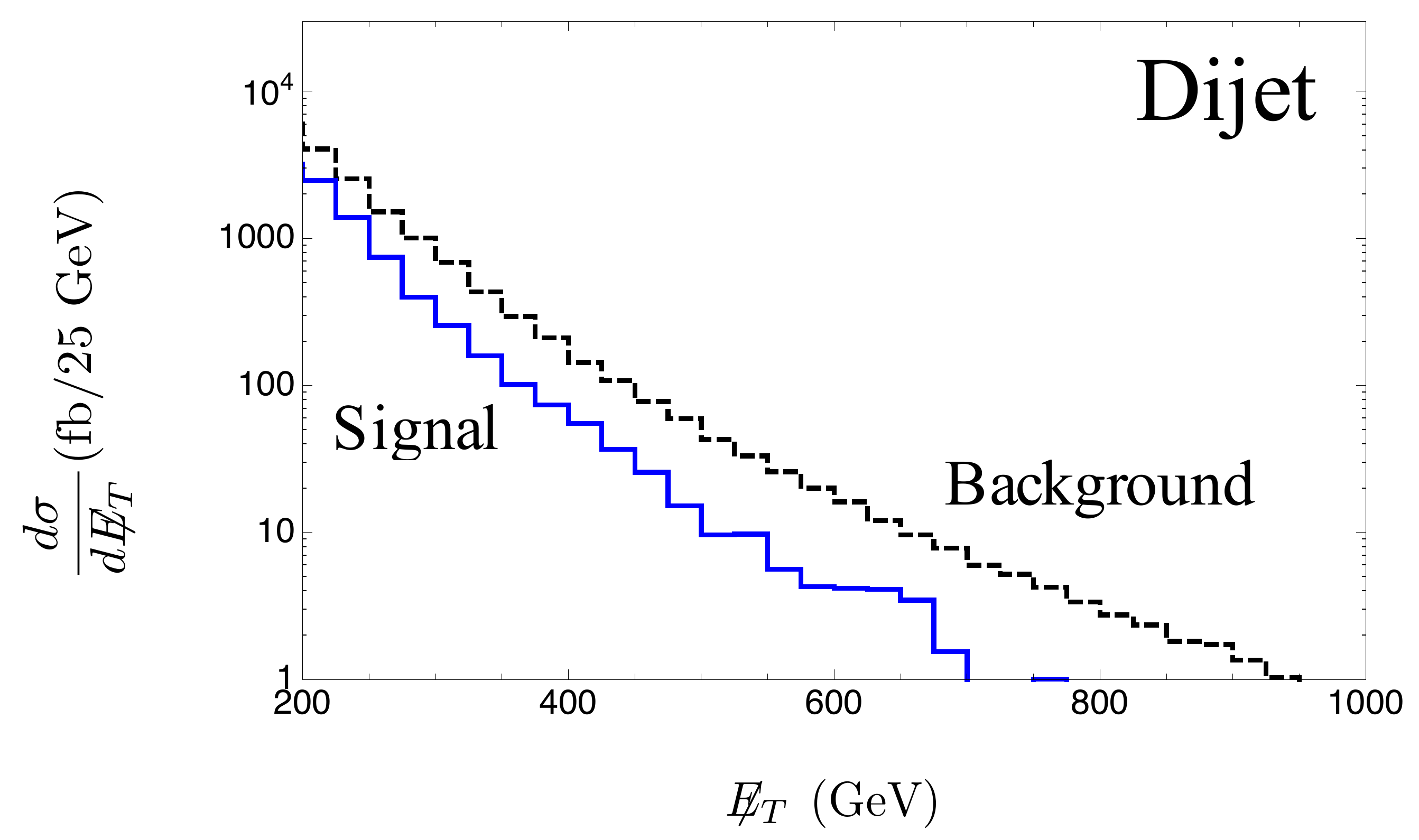} 
 \includegraphics[width=3in]{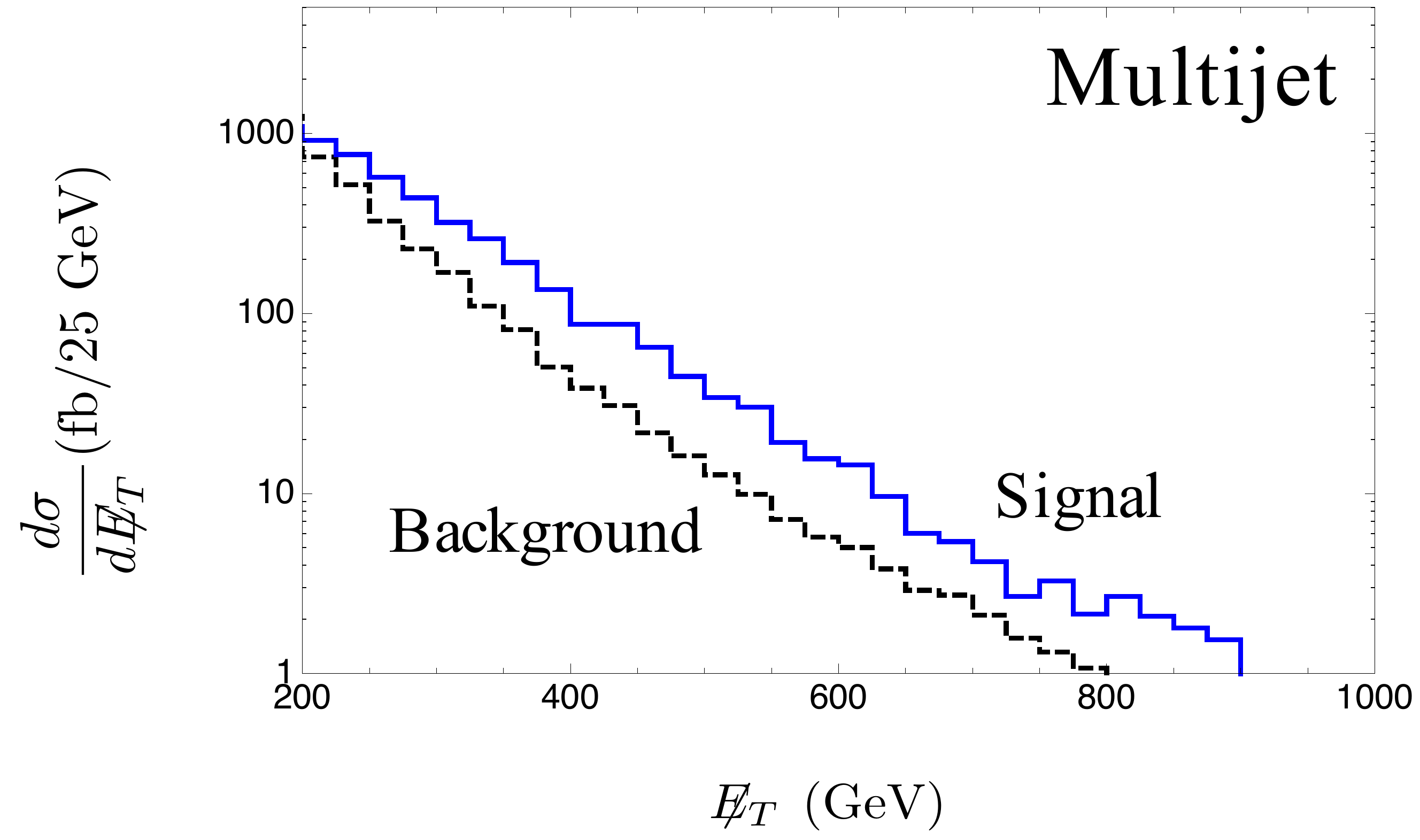}
  \caption{$\MET$ distributions for SM backgrounds (black) and a 250 GeV $\tilde{q}$ (blue) decaying to a 50 GeV LSP and a light flavored quark in dijet (left panel) and multijet (right panel)  channels.}
   \label{Fig: Squark Comparison}
 \end{center}
 \end{figure}

This study demonstrates that the exclusively multijet search strategy presented in Sec.~\ref{Sec: Strategy} and Sec.~\ref{Sec: Observables} is broadly applicable to many theories that do not have event topologies that appear dominantly as multijet events.  As a comparison between the reaches of the multijet and dijet channels, Fig.~\ref{MultivsDijet} shows discovery reach plots of both multijet and dijet and missing energy searches for pair-produced squarks decaying to a jet and the LSP, with $\sqrt{s}=14 \TeV$ and $\LL=1 \ifb$. Also shown is the reach of an inclusive dijet and $\MET$ search, where the $p_T$ of additional jets beyond the leading two is not vetoed. Moreover, increasing the $p_T$ requirement of the leading two jets in the dijet and $\MET$ search was also studied but found to be less sensitive than the criteria in Table~\ref{Tab: Classification}, particularly in the degenerate spectra region. 
Scalar color-triplets have a smaller production cross section than fermionic color-octets and results in lowering discovery potential at the LHC and is particularly difficult if the systematic errors are not reduced from 30\%.

\begin{figure}[htb]
\begin{center}
 \includegraphics[width=4.2in]{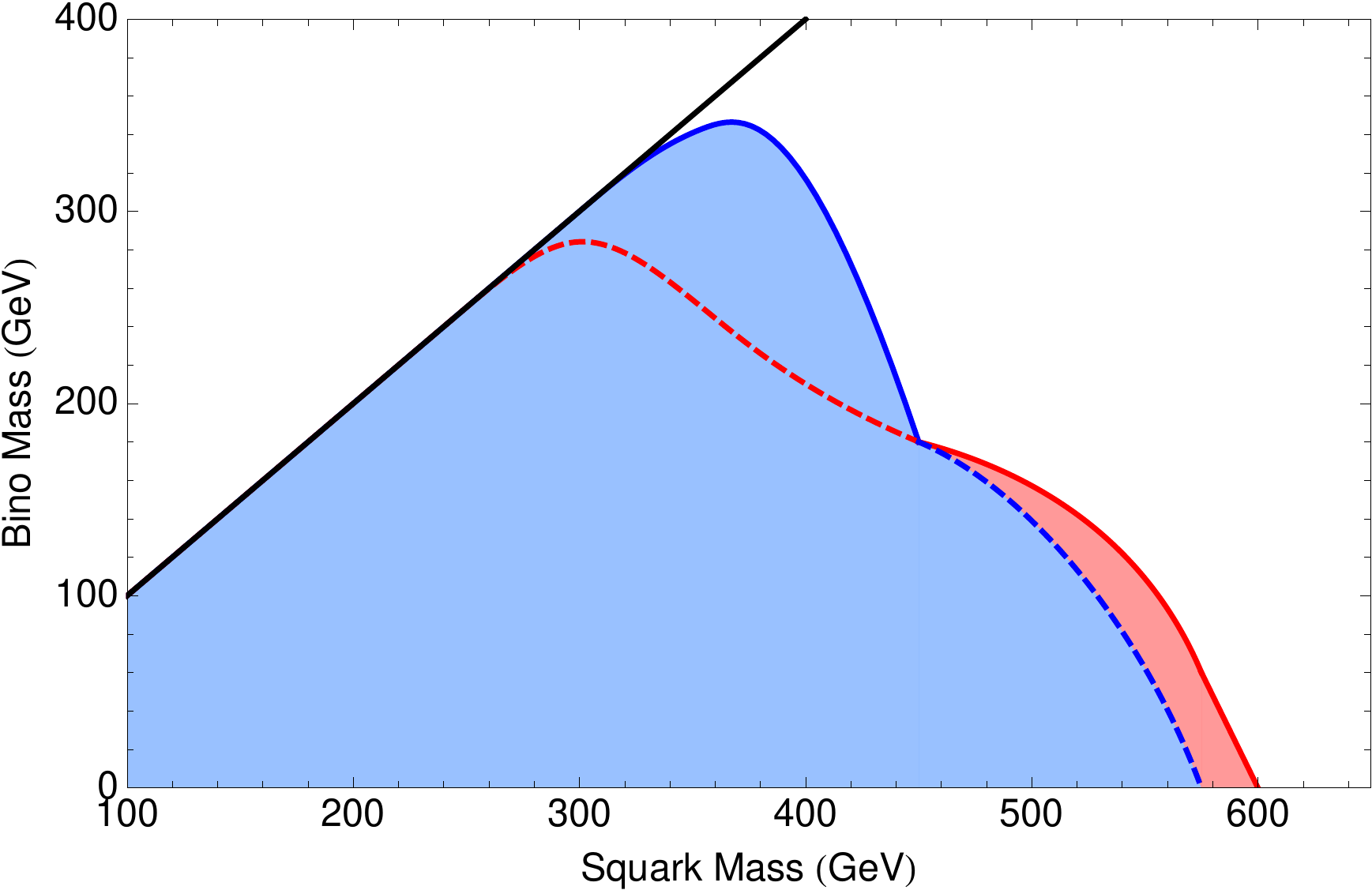}
 \caption{ Discovery reach at the LHC with $\sqrt{s}=14 \TeV$ and 1 fb$^{-1}$.  The light blue region shows  the reach of the multijet and $\MET$ high $\MET$ search, and the reach of the $\arts$ is shown in dashed red. }
 \label{MultivsDijet}
   \end{center}
 \end{figure}

 \subsection{Associated Squark Gluino Sensitivity}
As a final check, three benchmark points of squark-gluino associated production were computed:
\begin{eqnarray}
\nonumber
(m_{\tilde{g}}, m_{\tilde{q}}, m_{\chi^0}) & =&(400\GeV, 400 \GeV, 200 \GeV),\\  
\nonumber
&&(600 \GeV, 400\GeV,200 \GeV),\\ 
&& (250 \GeV, 400\GeV, 150\GeV).
\end{eqnarray}
In the first benchmark point, the gluino directly decays to two jets and the LSP, whereas the squark decays to a jet and the LSP. In the second case, the gluino decays to the squark and a SM quark; the squark, as in the previous case decays to a jet and the LSP.  In the final benchmark, point the squark decays to the gluino and a quark; the gluino, in turn, decays to two jets and the LSP.  Naively, the first and third benchmark points should be covered by trijet and missing energy searches, whereas the second point by a dijet and missing energy search. However, each of these  benchmark spectra is  separated from background by the multijet and missing energy searches depicted in Table~\ref{Tab: Different Searches}. 

\section{Discussion}
\label{Sec: Discussion}
 


Jets and missing energy searches are one of the most promising windows into new physics at the electroweak scale. The production of new colored particles comes with abundant energetic ISR and FSR jets. This article built upon this observation to propose a model-independent search strategy where nearly all classes of theories of beyond the SM containing colored particles which decay to jets and a neutral stable particle can be discovered in the $4^{+}j +\MET$ channel at $\sqrt{s}=14\TeV$. The approach is to perform a set of three searches that vary the amount of visible energy, missing energy, or $\arts$. 

This search strategy was applied to several classes of theories. The first considered pair-produced gluinos, with each gluino decaying to two jets and the LSP.  The second class of theories included pair produced gluinos that decayed in a cascade through an intermediate particle and finally to the LSP.    The third class of theories were pair-produced squarks  where each squark directly decayed to a jet and the LSP.  As a check, three benchmark spectra in the interpolating region of associated squark-gluino production were studied. 

The primary finding of this article is that incorporating parton shower matrix element matching to the signal increases the visibility of the signals by raising the missing energy in the event, particularly for heavy LSP masses.  This results in higher missing energy cuts  (greater than 400 GeV) being more effective at discovering new physics.  

Lower missing energy searches are not necessary until long cascades become normal.  If the LSP is a granddaughter of the produced particle, a 400 GeV missing energy cut is most effective at discovering the spectrum.   When the cascades become longer, so that the LSP is a great-granddaughter of the produced particle, the missing energy can be sufficiently degraded so that it is necessary to use lower missing energy cuts.
A typical 3-step cascade that has the LSP as  the great granddaughter can still be discovered effectively with high missing energy cuts, but those where there is a hierarchal spectrum of masses such as
\begin{eqnarray}
m_{\tilde{g}}: m_{\tilde{W}} : m_{\tilde{B}} : m_{\tilde{S}} = 8:4:2:1
\end{eqnarray}
where $\tilde{S}$ is a singlino that appears in many NMSSM models. 
The difficulty in using high $\MET$ requirements arises because the momentum of the final LSP carries so little of the original particle's  rest mass
energy or original momentum. Spectra such as
\begin{eqnarray}
m_{\tilde{g}}: m_{\tilde{W}} : m_{\tilde{B}} : m_{\tilde{S}} = 8:7:6:1\quad \text{ or }\quad m_{\tilde{g}}: m_{\tilde{W}} : m_{\tilde{B}} : m_{\tilde{S}} = 16:4:3:2
\end{eqnarray}
produce LSPs that still carry a reasonable fraction of the produced particle's momentum. 
In the first case,  most of the center of mass energy is distributed in the last decay.
In the second case, most is distributed in the first decay and then because the subsequent
 decays each have daughter particles that are non-relativistic, the momentum from the 
 original daughter particle is carried transferred to the granddaughter and then to the great granddaughter
 efficiently.

\noindent
\section*{ Acknowledgements}

We would like to thank  Johan Alwall, Mariangela Lisanti, and Jesse Thaler for helpful discussions.  We thank Daniele Alves, Jared Kaplan, and Mariangela Lisanti for reading early versions of the draft and providing incredibly valuable feedback.  EI and JGW thank the GGI institute for its hospitality in the later stages of this work. JGW is supported by the DOE under contract DE-AC03-76SF00515 and by the DOE's Outstanding Junior Investigator Award.

\providecommand{\href}[2]{#2}\begingroup\raggedright

\endgroup

\end{document}